\documentclass[prd,showpacs,preprintnumbers,amsmath,amssymb,superscriptaddress]{revtex4}
\usepackage{graphics}

\newcommand{\al}{\alpha}
\newcommand{\be}{\beta}
\newcommand{\ga}{\gamma}
\newcommand{\de}{\delta}
\newcommand{\s}{\sigma}
\newcommand{\pd}{\partial}

\newcommand{\brChr}[2]{\overline{\Gamma}^{#1}{}_{\!#2}}
\newcommand{\brg}{\bar{g}}
\newcommand{\brR}{\bar{R}}

\newcommand{\dd}{{\rm d}}

\newcommand{\eps}{\varepsilon}
\newcommand{\oem}{{\cal T}}
\newcommand{\brW}{\overline{\cal W}}

\newcommand{\blockmatrix}[4]{\left( \begin{array}{c|c} #1
& #2 \\ \hline #3 & #4 \end{array} \right)}

\begin{document}

\title{Cosmological tensor perturbations in the 
Randall--Sundrum model: \\ evolution in the near-brane limit}
\author{Richard A. Battye}
\affiliation{Jodrell Bank Observatory, Department of Physics and
Astronomy, University of Manchester, Macclesfield, Cheshire SK11
9DL, UK}
\author{Carsten van de Bruck}
\affiliation{Astrophysics Department, Oxford University, Keble
Road, Oxford OX1 3RH, UK}
\author{Andrew Mennim}
\affiliation{Department of Applied Mathematics and Theoretical
Physics, Centre for Mathematical Sciences, University of Cambridge,
Wilberforce Road , Cambridge CB3 OWA, UK}
\date{20 August 2003}
\preprint{DAMTP-2003-80}

\begin{abstract}
We discuss the evolution of cosmological tensor perturbations in the
RSII model. In Gaussian normal coordinates the wave equation is
non-separable, so we use the near-brane limit to perform the
separation and study the evolution of perturbations. Massive
excitations, which may also mix, decay outside the horizon which could
lead to some novel cosmological signatures.
\end{abstract}

\maketitle

\section{Introduction}
The brane-world idea, according to which the observable universe 
is a hypersurface (brane) embedded in a higher-dimensional spacetime (bulk), 
has attracted the attention of particle physicists and cosmologists in
recent years --- for reviews see~\cite{review1,review2,review3,review4}. 
The idea is motivated from developments in string theory and
M--theory, although many of the models which have been 
developed are phenomenological.
Two such models were constructed by Randall and
Sundrum~\cite{RS1,RS2}, in which the bulk spacetime is a
five-dimensional Anti-de Sitter (AdS) spacetime with a small
length-scale ($\sim 1\,$mm or less). 
In this article, we will be considering the second~\cite{RS2} of these
models which has only one brane with a large, positive, bare tension
to balance the curvature of the AdS bulk.
The cosmological background solutions in the bulk spacetime were
calculated~\cite{BDL2,FTW1,FTW2} shortly after the model was proposed.

Cosmological perturbations around this background, however, have resisted
a solution in the face of a considerable research
effort~\cite{perturbationsfirst,kodama,roypertur,langlispertur,vdbruckpertur,koyama,royandchris,langloisnew,LMW,malik2,nathalie,vdb2,vdb3,malik3,soda2,bernhard1,bernhard2,LMSW,deffayet,KolbPert,perturbationslast}.
(For a review on cosmological perturbations in brane-worlds see \cite{nathaliereview}.)
Although diverse formalisms for cosmological perturbations in brane models exist, 
the evolution of perturbations during the different cosmological 
epochs is not yet fully understood. The problem is that the bulk equations 
are partial differential equations which are subject to 
spacetime dependent boundary conditions on the brane. Therefore, the full 
problem can cannot be reduced to a four-dimensional effective problem.
In addition, there is a large difference in scales: the AdS
length-scale is of the order of $1\,$mm or less, whereas, the perturbations 
have to be evolved over many orders of magnitudes of conformal time,
a ratio in scales of about $10^{26}$.
Because of this, the wave equation for the evolution of the
perturbations is difficult to solve numerically, at least
by brute-force numerical integration:
the spatial length scale of $1\,$mm means that we would have to
use a corresponding time scale as the time-step, meaning the
code would have to be run for the order of $10^{26}$ time steps!

In order to test the brane-world idea against cosmological 
data, it is extremely important to understand the evolution of 
perturbations. One of the most powerful discriminating tools between cosmological
models is the power spectrum of the cosmic microwave background (CMB)
anisotropies, which is, effectively, a snapshot of the perturbations in the universe
when it was about one thousandth of its current size.
At about this time in the history of the universe, two related events
occur: atomic recombination and last scattering of photons.
Before atomic recombination, the temperature of the universe is so high
that protons and electrons form a plasma, but during
recombination they come together to form atomic hydrogen.
At approximately the same time, the mean free path of the photons due Thomson scattering in the
plasma increases rapidly from being very small to very large (effectively,
from zero to infinity) meaning that most of the photons arriving at
the earth from the cosmological plasma will not have scattered
subsequently. Once foreground effects due to our galaxy and external galaxy 
clusters have been removed, the electromagnetic radiation seen from earth is an image
of this surface of last-scattering, or, more precisely, the
intersection of it with our past light-cone. 
However, because of perturbations in the early universe, this will
occur at slightly earlier or later times at different points of space,
so the CMB radiation will be slightly brighter or dimmer at
different points on the sky.
The upshot of this is that, by measuring the temperature fluctuations
in the CMB, we can deduce the perturbations in the metric on the
surface of last scattering.

The details of the perturbations in the metric at the time of last-scattering 
and their subsequent evolution depend on the theory of gravity. In brane-world 
models, Einstein's equation becomes modified, at least at high
energies, and it is plausible that these effects can propagate into the
regime relevant to the creation of CMB anisotropies. Clearly, the
evolution of cosmological perturbations in this model is of
significant interest. 

The modifications in the Randall--Sundrum (RS) model 
are twofold: first, in Einstein's equation a term quadratic 
in the energy-momentum tensor appears. In the context of cosmology, this implies that at high 
energies the Hubble parameter on the brane is proportional the energy density of matter $\rho$ 
and not proportional to $\sqrt{\rho}$, as in General Relativity. However, in 
the RS model one can show that this term is negligible at 
low energies, i.e., at energies much smaller than the brane tension. 
The other correction which appears in Einstein's field equation is the 
five-dimensional Weyl tensor projected onto the brane. This term encodes 
the influence of the bulk gravitational field on the dynamics of the brane. 
In the case of a homogeneous and isotropic brane this term vanishes in the 
case of an AdS bulk, but is non-zero if the bulk contains a black hole 
(Schwarzschild--AdS). From the point of view of 4D cosmology, this term behaves like a smooth energy component with equation of state $P=\rho/3$ and was 
therefore dubbed ``dark radiation''\cite{BDL2}. Its response to perturbations is somewhat different~\cite{langloisnew}. 

The purpose of this article is to study the evolution of tensor perturbations 
in the RS brane-world. Although these perturbations are easier 
to understand than the observationally interesting case of 
scalar perturbations, we will see that their evolution is already
very complicated.
There are several coordinates system which one might use to tackle the
problem.
In the Gaussian normal (GN) coordinate system the brane is at rest but the
bulk metric components have a complicated time-dependence:  the brane
boundary condition are easy to impose, but the solution of the
corresponding wave equation is much more subtle since it is non-separable.
Alternatively, one can formulate the problem in a coordinate system
where the brane moves, but the bulk spacetime is manifestly static:
one can then solve the wave equation in the bulk, but the boundary
condition on the brane is difficult to impose.
In this paper we will discuss solutions in the near-brane limit of
the GN coordinate system.  These solutions are the most
direct analogue of the well studied mode functions of Minkowski
space. Some of the material presented here first appeared in
\cite{andyproc} and \cite{andythesis}. The fact that the equation of
motion is not separable is likely to lead to mode mixing and this will
be discussed in future work.

The paper is organized as follows: In Section~\ref{Overview} we discuss the 
different scales involved and discuss different ways to 
attack the problem. In Section~\ref{Setup} we present the setup and derive the 
bulk wave equation and the brane boundary conditions. We also discuss the
issue of initial conditions. In Section~\ref{bases} we present the near-brane solutions 
in the different cosmological epochs. In Section~\ref{secIC} we 
discuss the problem of initial conditions. 
We present our conclusions in Section~\ref{conc}.

\section{Overview}
\label{Overview}

For simplicity, we will only consider the second RS model~\cite{RS2}
which has a single brane of constant positive bare tension in an
Anti-de Sitter (AdS) bulk which has a reflection symmetry between the
two sides of the brane.
Furthermore, we will restrict our attention to the case of a spatially
flat universe, which is observationally favoured.
The bulk Einstein equation for this model is then
\begin{equation}
\label{backgroundEinstein}
R_{\mu\nu}=\frac{2}{3}\Lambda g_{\mu\nu} \,.
\end{equation}
The Gauss relation gives the four-dimensional Ricci tensor as
\begin{equation}
\label{4dRicci}
\brR_{\mu\nu}=\frac{\Lambda}{2}\brg_{\mu\nu}+KK_{\mu\nu}-
{K_\mu}^\rho K_{\nu\rho}-\brW_{\mu\nu} \,,
\end{equation}
where the extrinsic curvature tensor of the brane is defined as
\begin{equation}
K_{\mu\nu}=-\brg_\mu{}^{\!\rho}\nabla_\rho n_\nu \,,
\end{equation}
with $n_\nu$ being the unit covector normal to the brane, and 
\begin{equation}
\brW_{\mu\nu}=C_{\mu\rho\nu\s} n^\rho n^\s\,,
\end{equation}
is the contribution from the Weyl tensor of the bulk
spacetime~\footnote{We are using the conventions of Misner, Thorne and
Wheeler~\cite{MTW} throughout. In particular, many authors use the
opposite sign convention for the extrinsic curvature tensor.}.
The junction conditions~\cite{MTW} become
\begin{equation}
\label{Israel}
K_{\al\be}=\frac{\kappa}{6}\Big\{3\oem_{\al\be}+
(\oem-\lambda)\brg_{\al\be}\Big\} \,,
\end{equation}
whilst the definition of $K_{\al\be}$, combined with the bulk Einstein
equations (\ref{backgroundEinstein}), yields
\begin{eqnarray}
\label{Kevolve}
\pounds_n K_{\al\be}&=&\brW_{\al\be}+\frac{\Lambda}{6}-
K_{\al\ga}K_\be{}^{\!\ga}\,,\\
\pounds_n\brg_{\al\be}&=&-2K_{\al\be} \,.
\label{gevolve} 
\end{eqnarray}
We will study perturbations by linearizing these equations about
the cosmological background solution.
For simplicity, we will focus on the case of a spatially flat
universe, i.e., where the surfaces of isometry are flat planes,
and consider only the case of a pure AdS bulk where $\brW_{\al\be}=0$
for the background.

\subsection{The variety of coordinates and gauges}

There are a variety of formalisms available for studying this problem.
In this section, we explore some of these possibilities and discuss
their various merits and drawbacks.
There is no clear favourite, but some gauges are better for studying
certain aspects of the  problem.

There are two different natural choices of coordinates in which
it is convenient to express the background, in addition to the
possibility of a coordinate-independent formalism.
One natural choice is the Gaussian normal (GN) coordinate system.
In these coordinates the metric for the background (non-perturbed) spacetime has the form
\begin{equation}
\dd s^2=-n(\tau,\zeta)^2\,\dd\tau^2+a(\tau,\zeta)^2\de_{ij}
\, \dd x^i \dd x^j+\dd \zeta^2 \,.
\end{equation}
The main advantage of this coordinate system is that the brane remains
at a fixed value of one of the coordinates, meaning that imposing 
the boundary condition due to the perturbed junction conditions is simple.
The main disadvantage is that the metric components $a$ and $n$ have
complicated functional forms and, as a consequence, the equations of
motion are not separable.

The other natural coordinate systems are ones where the static nature
of the background is manifest.
Another way of writing the line element for a flat cosmology in an AdS
bulk is
\begin{equation}
\label{confminkds}
\dd s^2=\frac{l^2}{z^2} \eta_{\al\be}\, \dd x^\al \dd x^\be
=\frac{l^2}{z^2}\Big(-\dd t^2+\de_{ij}\,\dd x^i \dd x^j
+\dd z^2\Big) \,,
\end{equation}
where the spacetime is manifestly conformally flat.
(Here, $l$ is the AdS length-scale.)
This coordinate system makes the five-dimensional linearized Einstein
equations simple.
The price paid for this simplicity is that the brane is no longer at a
fixed coordinate value, but has locus given by $z=l/a$, making the
boundary condition much more difficult to impose. 

In addition to the various coordinate systems for expressing the
background, there are also various gauge choices for expressing the
perturbations, as well as gauge-independent formalisms such as those
used in~\cite{perturbationsfirst,kodama,bernhard1,bernhard2,perturbationslast}.
The gauge we will use is the GN gauge, which arises by requiring that
both the perturbed and unperturbed metrics have GN form, so that the
perturbed line element has the form
\begin{equation}
\dd s^2=-n^2(1+\phi)\,\dd\tau^2+2anb_i\,\dd\tau\,\dd x^i+a^2\,(\de_{ij}+h_{ij})
\,\dd x^i\dd x^j+\dd\zeta^2\,.
\end{equation}
When using a GN background, this is a very natural choice and has the
advantage of making it trivial to extract the perturbation of the
brane metric, $\brg_{\mu\nu}$.
Note that this does not completely fix the gauge since it it possible
to impose additional requirement on the values taken by certain components
of 
perturbation variable on the brane.
For example, we could impose $\phi=b_i=0$ on the brane, which would
make the perturbation of the brane metric that of the synchronous
gauge, a gauge often used in the treatment of perturbations in the
standard, four-dimensional cosmology \cite{MaBert}.

If we chose to work in the conformally Minkowski coordinates of
(\ref{confminkds}), we would naturally write the perturbed metric as
\begin{equation}
\dd s^2=\frac{l^2}{z^2}\Big( \eta_{\al\be}+h_{\al\be} \Big)
\dd x^\al \dd x^\be\,,
\end{equation}
which would allow us to make various gauge choices.
For example, \cite{nathalie} used the transverse-traceless (TT) gauge where
$h_{\al\be}$ satisfies
\begin{equation}
\eta^{\al\be}h_{\al\be}=0 \,,\qquad \pd_\al h^\al{}_{\!\be}=0\,.
\end{equation}
Late in the history of the universe, the motion of the brane in this
coordinate system is slow, so that the time coordinate, $t$, is
almost the same as conformal time on the brane and so the TT gauge is 
approximately a synchronous gauge.
This setup has the advantage that the resulting perturbed Einstein
equations exactly soluble.
However, in this gauge the position of the brane is not on the same
locus as for the background but is displaced, an effect which has
been dubbed ``brane-bending'' in the literature.

\subsection{Physical scales in the problem}

We will be considering the evolution of perturbations by taking the
Fourier transform in the three spatial directions of the surfaces of
isotropy and then evolving each Fourier mode.
Thus, each mode has associated with it a length-scale given by
$k^{-1}$, where $k$ is the magnitude of the Fourier transform
variable.
Physically, we interpret $k$ as the comoving momentum of the mode and
$k^{-1}$ as the comoving wavelength.
There is another length-scale, $H^{-1}$, given by the time dependent 
expansion rate of the universe,  equivalent to
the Hubble radius of the Universe.
In the standard cosmology, these are the only two length-scales,
giving rise to a single dimensionless parameter, $kH^{-1}$, which
depends on time.
In the language of cosmological perturbation theory, this parameter
tells us whether the mode is outside or inside the Hubble horizon.

For perturbations in the RS model, there is another length-scale, namely
the AdS length-scale, $l$.
Thus, we have three dimensionless parameters: $kH^{-1}$, $lH$ and
$kl$, two of which are independent.
The first two of these depend on time: $kH^{-1}$ becomes larger at
later times, whereas, $lH$ is smaller at later times. 
Both will, of course, remain approximately constant during a period of
inflation.
As in the four-dimensional case, $kH^{-1}$ gives the scale of the mode
relative to the horizon size.
Typically, the AdS length-scale, $l$ will be less than $1\,$mm whereas
the length-scale of the perturbations $k^{-1}$ will be greater than
$1\,$Mpc, so $kl$ will be an extremely small number.
Except in the very early universe, $lH$ will also be small since it is
the Hubble parameter measured in time units corresponding to $1\,$mm.

\section{Setup}
\label{Setup}

As already discussed, there are a variety of formalisms, each with its
own merits.
Here, we shall use a GN coordinate system and gauge, so that the
perturbed line element takes the form
\begin{equation}
\label{pertds}
\dd s^2=-n^2(1+\phi)\,\dd\tau^2+2anb_i\,\dd\tau\,\dd x^i+a^2\,(\de_{ij}+h_{ij})
\,\dd x^i\dd x^j+\dd\zeta^2\,.
\end{equation}
The brane is located at $\zeta=0$.  Note that our choice does not
completely fix the gauge as, for example, we can set $\phi$ and $b_i$
to be zero on the brane, thereby having the brane metric in
synchronous gauge.
As a shorthand, we will denote derivatives with respect to $\tau$ and
$\zeta$ by dots and dashes respectively. Note that $\phi$ and $b_{i}$ are irrelevant for the tensor perturbations studied here.

The observed matter, $\oem_{\al\be}$, will comprise a background part
into a background part
\begin{equation}
\oem^\al{}_{\!\be}=
\blockmatrix{-\rho}{0}{0}{p{\delta^i}_j} \,,
\end{equation}
and a perturbation
\begin{equation}
\label{empert}
\de\oem^\al{}_{\!\be}=
\blockmatrix{-\delta \rho}{an^{-1}(\rho+p)(v_j+b_j)}{-a^{-1}n(\rho+p)v^i}
{\delta p\,\de^i{}_j+\Sigma^i{}_j} \,,
\end{equation}
with ${\Sigma^i}_i=0$.
This is similar to the formalism used in~\cite{LMSW}.
It has the great advantage that the brane metric in synchronous gauge
can easily be read-off and standard CMB computer code (such as
{\tt CMBFAST}~\cite{MaBert}) can be applied to solve the Boltzmann equations
and determine the power spectrum of matter and CMB anisotropies.

For the RS model, the bulk spacetime is pure AdS, which has a length-scale,
$l$, related to the cosmological constant by
\begin{equation}
l^2=-\frac{6}{\Lambda}\,.
\end{equation}
The form of the functions $a(\tau,\zeta)$ and $n(\tau,\zeta)$ was found in
\cite{BDL2} to be
\begin{eqnarray}
\label{sfa}
a(\tau,\zeta)&=&a(\tau,0)\Bigg[\frac{1}{2}\left(1+\frac{U(\tau)^2}{6\Lambda}\right)
+\frac{1}{2}\left(1-\frac{U(\tau)^2}{6\Lambda}\right)\cosh(2\zeta/l)
-\sqrt{\frac{-U(\tau)^2}{6\Lambda}}\sinh(2\zeta/l)\Bigg]^{1/2}\,,\\
n(\tau,\zeta)&=&n(\tau,\zeta)=\frac{\dot{a}(\tau,\zeta)a(\tau,0)}{\dot{a}(\tau,0)}\,,
\label{sfn}
\end{eqnarray}
where $U(\tau)=\lambda+\rho(\tau)$ is the total energy density supported on the
brane, i.e., the bare tension, $\lambda$, plus energy density of observed matter,
$\rho(\tau)$\footnote{Note that the solutions given in \cite{BDL2} are for the, more general, Schwarzschild-AdS spacetime and so look more complicated.}.
The metric component $n(\tau,\zeta)$ is chosen so that $\tau$ corresponds to conformal
time on the brane.
We will also tune the bare tension against the bulk cosmological constant
so as to make the effective brane cosmological constant zero, for
which we need to take $\lambda=3/(4\pi Gl^2)$.
With this we get
\begin{equation}
\label{aexpr}a(\tau,\zeta)
=a(\tau,0)\left[e^{-\zeta/l}-\frac{\rho}{\lambda}\sinh(\zeta/l)\right] \,.
\end{equation}

For convenience, we list the non-zero Christoffel symbols for this background,
which are
\begin{equation}
\label{backgroundChr}
\brChr{0}{00}=\frac{\dot{n}}{n}\,,\quad
\brChr{i}{0j}=\frac{\dot{a}}{a}\,{\delta^i}_j\,,\quad
\brChr{0}{ij}=\frac{a\dot{a}}{n^2}\,\delta_{ij}\,,
\end{equation}
and the non-zero components of the Riemann tensor, which are
\begin{equation}
\brR^{0i}{}_{0j}=\frac{n\ddot{a}-\dot{n}\dot{a}}{an^3}\,{\delta^i}_j\,,\quad
\brR^{ij}{}_{kl}=2\frac{\dot{a}^2}{a^2n^2}\,{\delta^{[i}}_{[k}{\delta^{j]}}_{l]}\,.
\end{equation}
The bulk Einstein equations give the following
relations between the metric components
\begin{eqnarray}
-\frac{2}{3}\Lambda&=&\frac{n''}{n} - 3\frac{\ddot{a}}{an^2}
+ 3\frac{\dot{a}\dot{n}}{an^3} + 3\frac{a'n'}{an}\,,\\
-\frac{2}{3}\Lambda&=&\frac{n''}{n} + 3\frac{a''}{a}\,,\\
-\frac{2}{3}\Lambda&=&-\frac{\ddot{a}}{an^2}+\frac{\dot{a}\dot{n}}{an^3}
+\frac{a'n'}{an}+\frac{a''}{a}-2\frac{\dot{a}^2}{a^2n^2}
+2\frac{{a'}^2}{a^2}\,,\\
0&=&\frac{\dot{a}'}{a}+\frac{\dot{a}n'}{an}\,.
\end{eqnarray}
The brane, at $\zeta=0$ has extrinsic curvature given in
Gaussian normal coordinates by
\begin{equation}
K_{\al\be}=-\frac{1}{2}\frac{\dd}{\dd\zeta}\brg_{\al\be} \,,
\end{equation}
which has components
\begin{equation}
\label{backgroundK}
K_{00}=nn'\,,\quad K_{0i}=0\,,\quad
K_{ij}=-aa'\de_{ij}\,, 
\end{equation}
from which we can deduce that 
\begin{equation}
K=-\left(3\frac{a'}{a}+\frac{n'}{n}\right) \,.
\end{equation}
Note that some authors use the opposite sign convention for the
extrinsic curvature.

\subsection{Equations of motion and brane boundary conditions}

Since we have chosen GN gauge, where the perturbed metric is in GN
form, it is natural to consider the perturbation of the Gauss relation
(\ref{4dRicci}) which is
\begin{equation}
\label{GaussPert}
\de\brR_{\mu\nu}=\frac{\Lambda}{2}\de\brg_{\mu\nu}+
\de K K_{\mu\nu}+K\de K_{\mu\nu}+K_\mu{}^\rho K_\nu{}^\s
\de\brg_{\rho\s}-2K_{(\mu}{}^\rho\de K_{\nu)\rho}-\de\brW_{\mu\nu}\,,
\end{equation}
and perturbing (\ref{Kevolve}) and (\ref{gevolve}) gives
\begin{eqnarray}
\label{diffKpert}
\frac{\dd}{\dd\zeta}\de K_{\al\be}&=&2K_{(\al}{}^\rho\de K_{\be)\rho}
-K_\al{}^\rho K_\be{}^\s\de\brg_{\rho\s}-\frac{\Lambda}{6}
\de\brg_{\al\be}-\de\brW_{\al\be} \\
\frac{\dd}{\dd\zeta}\de\brg_{\al\be}&=&2\de K_{\al\be}
\label{diffgpert}
\end{eqnarray}
substituting these into (\ref{GaussPert}) and equating with the
perturbations of the Ricci tensor calculated from the metric gives
the evolution equations for the metric perturbations.
We will evaluate these evolution equations in the special case of
tensor perturbations.
Perturbing the Israel conditions of (\ref{Israel}) gives
\begin{eqnarray}
\de K_{00}&=&-\frac{\kappa}{6}n^2\Big\{2\de\rho
+3\de p+(2\rho+3p)\phi\Big\}\,,\\
\de K_{0i}&=&\frac{\kappa}{6}an\Big\{3(\rho+p)v_i+
(2\rho+3p-\lambda)b_i\Big\}\,,\\
\de K_{ij}&=&-\frac{\kappa}{6}a^2\Big\{\de\rho\de_{ij}+
(\rho+\lambda)h_{ij}+3\Sigma_{ij}\Big\}\,,
\label{Israelij}
\end{eqnarray}
which will provide the boundary conditions we need at the brane.

\subsection{Initial conditions and boundary conditions at the horizon}

Since we are aiming to evolve the equations of motion over the
semi-infinite interval $0 \le \zeta \le \infty$ it would appear that a
another boundary condition is not necessary.
However, the coordinates used do not span the universal covering set
of the AdS spacetime (see \cite{HawkingEllis} for more details on the
universal covering set) so we could expect that gravitational waves
could cross the AdS horizon.
The choice of boundary condition on this horizon is somewhat arbitrary
since we can have no knowledge of the other side; this being a philosophical shortcoming of the Randall--Sundrum model.
A reasonable, and popular, choice is to have no incoming radiation at
the horizon, as advocated in~\cite{Rubakov}.

\begin{figure}
\centerline{\scalebox{0.7}{\includegraphics{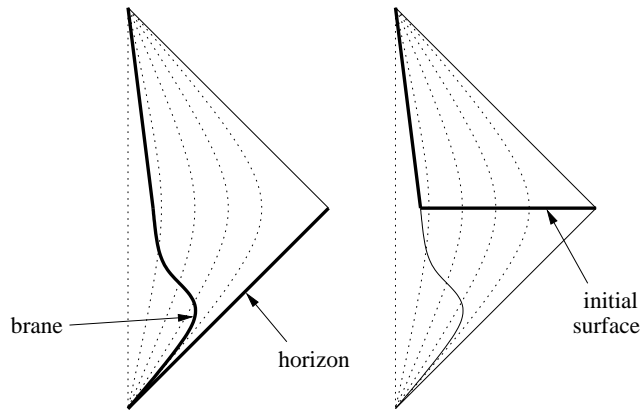}}}
\caption{Specification of boundary data: on the left, at the brane and
the horizon; on the right, at the brane and on some initial surface.
The bold lines show where boundary date is specified.
The dotted lines are for guidance and indicate surfaces with a
Minkowski four-metric.}
\label{CPdiagram}
\end{figure}

However, it is not possible to impose this boundary condition in
addition to arbitrary initial conditions on some time-slice.
So we can impose boundary data in two different ways:
\begin{enumerate}
\item Brane boundary condition and a condition at horizon.
\item Brane boundary condition and initial conditions at some time.
\end{enumerate}
The second of these is more in keeping with the orthodox approach to
cosmological perturbation theory. However, it requires knowledge of
what created in the first place.
These are not as different as they might seem, we can gain a better
understanding of how they are related, and of when it is necessary to
specify boundary data at the horizon by considering Carter--Penrose
conformal diagrams of the spacetime, see Fig.~\ref{CPdiagram}.
It is apparent from the diagrams that specifying data on the horizon
is a limiting case of specifying data on a timelike initial surface.

To illustrate this further, it is useful to consider the solution of
the ordinary, wave equation in one spatial dimension,
\begin{equation}
\frac{\pd^2\phi}{\pd x^2}=\frac{\pd^2\phi}{\pd t^2}\,,
\end{equation}
with a boundary condition $\pd\phi/\pd x=0$ at $x=0$.
The spatial part of the mode solutions are then of the form
$\cos(\omega x)$, multiplied by a time dependent part of the same
frequency, $\omega$.
If we then impose a no incoming radiation condition, the solution is
completely determined, and is constant, so it is not possible to
impose initial data in addition.
This can be seen more clearly by writing the general solution in the
form $f(x-t)+g(x+t)$.  The no radiation condition would force us to
take $g\equiv0$ and the brane boundary condition necessitates that
$f\equiv{\rm constant}$.

For studying the initial perturbation spectrum created in the inflationary
era, it is useful to use the no-radiation boundary condition, as was
done in~\cite{Rubakov}.
Once we have the perturbations at the end of inflation, we can evolve
them forward through the radiation eras, for which we will have
initial data on a time-slice, and then through the matter era.

\subsection{Tensor modes}

The perturbations in the four-dimensional part of the metric naturally
split into scalar, vector and tensor parts with respect to the Fourier
transform variable, $k^i$.
To linear order in perturbation theory, they do not mix.
Of these the scalar perturbations are the most important since these
contain the underdensities and overdensities giving rise to
large-scale structure and so necessarily exist. 
As we have already suggested, it is much easier to solve the equations for the tensor
perturbations.
We have formulated the equations of motion for all three of these but
will only provide solutions for the tensor modes.
Some of the lessons learned should assist in tackling the more
difficult problem of the scalar modes.

The variation of the purely spatial part of the Ricci tensor is given by
$\delta\brR_{ij}=\delta\brR^0{}_{i0j}+\delta\brR^k{}_{ikj}$,
so the tensor part is given by
\begin{equation}
\label{Rtens1}
\delta\brR^T_{ij}=\frac{a^2}{2n^2}\ddot{h}^T_{ij}+\frac{a^2}{2n^2}
\left(3\frac{\dot{a}}{a}-\frac{\dot{n}}{n}\right)\dot{h}^T_{ij}+
\frac{a^2}{n^2}\left(\frac{\ddot{a}}{a}-\frac{\dot{a}\dot{n}}{an}+
2\frac{\dot{a}^2}{a^2}\right)h^T_{ij}+\frac{k^2}{2}h^T_{ij}\,,
\end{equation}
where the superscript $T$ represents either of the two tensor modes,
which will evolve separately, to linear order, and will both obey the
same equation of motion.
Equations (\ref{diffKpert}) and (\ref{diffgpert}) give us
\begin{eqnarray}
K^T_{ij}&=&\frac{1}{2}\frac{\dd}{\dd\zeta}\left(a^2h^T_{ij}\right) \,, \\
E^T_{ij}&=&\frac{a'}{a}\frac{\dd}{\dd\zeta}\left(a^2h^T_{ij}\right)
-{a'}^2h^T_{ij}-\frac{\Lambda}{6}a^2h^T_{ij}
-\frac{1}{2}\frac{\dd^2}{\dd\zeta^2}\left(a^2h^T_{ij}\right)\,.
\end{eqnarray}
Substituting these into the tensor part of (\ref{GaussPert}) gives
\begin{equation}
\label{Rtens2}
\delta\brR^T_{ij}=\left(\frac{2\Lambda}3+2\frac{{a'}^2}{a^2}\right)a^2h^T_{ij}
+\frac{1}{2}\left(\frac{n'}{n}-\frac{a'}{a}\right)\frac{\dd}{\dd\zeta}\left(a^2h^T_{ij}\right)
+\frac{1}{2}\frac{\dd^2}{\dd\zeta^2}\left(a^2h^T_{ij}\right)\,.
\end{equation}

The purely spatial part of the Einstein equation
(\ref{backgroundEinstein}) evaluated for the 
background metric gives
\begin{equation}
\frac{2}{3}\Lambda=\frac{\ddot{a}}{an^2}+2\frac{\dot{a}^2}{a^2n^2}-
\frac{\dot{a}\dot{n}}{an^3}-\frac{a''}{a}-\frac{a'n'}{an}-
2\frac{{a'}^2}{a^2}
\end{equation}
using this and equating (\ref{Rtens1}) and (\ref{Rtens2}) gives
\begin{equation}
\label{tensorpertevol}
\ddot{h}^T_{ij}+\left(3\frac{\dot{a}}{a}-\frac{\dot{n}}{n}\right)\dot{h}^T_{ij}
+k^2\frac{n^2}{a^2}h^T_{ij}-n^2\left(3\frac{a'}{a}
+\frac{n'}{n}\right)\frac{\dd}{\dd\zeta}h^T_{ij}
-n^2\frac{\dd^2}{\dd\zeta^2}h^T_{ij}=0 \,,
\end{equation}
in agreement with the result in~\cite{LMW}.

The boundary condition on $h^T_{ij}$ at the brane is given by the
perturbation of the purely spatial part of the junction condition
(\ref{Israelij}) which becomes
\begin{equation}
\label{tensorjunction}
\left.{h^T_{ij}}'\right|_{z=0} = -\kappa\Sigma_{ij}^T\,.
\end{equation}
The tensor part can be written in terms of two polarization tensors
$e^+_{ij}$ and $e^\times_{ij}$ as
\begin{equation}
h^T_{ij}=h^+ e^+_{ij} + h^\times e^\times_{ij}\,,
\end{equation}
so, in subsequent equations, we will simply write $h$ to
represent either $h^+$ or $h^\times$.

For most of the history of the universe, there is effectively 
no matter source for
the tensor modes so this condition will reduce to $h'=0$ on the
brane.  Only when the CMB photons develop a quadrupole moment in the
late universe is this assumption no longer valid.

The wave equation (\ref{tensorpertevol}) is not separable and is difficult
to solve in general, due to the complicated form
of the scale factors $a$ and $n$, see (\ref{aexpr}).
We will first study two cases where the equations can be solved
exactly: the case of a Minkowski brane and the case of a de Sitter
brane.
The first of these was the case considered by Randall and Sundrum in
their original paper~\cite{RS2}; the second is very useful in
cosmology for modeling an inflationary era in the early universe~\cite{LMW,Rubakov}.

\section{Basis solutions}
\label{bases}

\subsection{Near-brane limit}

Our main interest is in cases which  correspond to different cosmological eras.
In an inflationary model, the universe will undergo a phase of
accelerated expansion where the brane metric will be approximately de
Sitter.
After a transition period, there will be an era where the matter
content of the universe is dominated by a radiation fluid component,
having equation of state $p=\rho/3$.
There will then be another transition to evolution dominated by
pressureless matter.
Shortly after this transition, photons decouple from the
cosmological plasma, producing the CMB radiation.
Since the eventual aim of studying perturbations is to determine
theoretical CMB power spectra for RS models and compare these to the
observed CMB spectrum, we should try to solve the problem in each of
these eras.

Unfortunately, the equation of motion is not separable in the
radiation and matter dominated eras so we will approximate the
solution by one valid near the brane, ignoring the $\sinh(\zeta/l)$
term in (\ref{aexpr}).
The range of $\zeta$ for which the near-brane limit 
 is a good approximation depends on
the density of the cosmological fluid.
The value of $\zeta$ for which $a(\tau,\zeta)=0$ is given by
\begin{equation}
\label{goodapprox}
\zeta=\frac{l}{2}\log\left(1+\frac{2\lambda}{\rho}\right)\,,
\end{equation}
which will give us a qualitative idea of the domain of validity of
such a solution.
Since $\rho$ will decrease with time in the radiation and matter eras,
this domain of validity will get wider later in the history of the
universe. The near-brane limit can also be thought of in some sense as a low energy limit, but the fact that it is an approximation suggests that discussion of a low-energy effective theory needs to be careful thought out.

Let us see where the locus of $a=0$ intersects the past light cone of
a point on the brane today since this will correspond to the limit of causal validity of the near-brane limit.
It is simpler to work in proper time on the brane, $t$, so that a null
geodesic satisfies
\begin{equation}
\label{nullDE}
\frac{\dd\zeta}{\dd t}=\pm n(t,\zeta) \approx \pm e^{-\zeta/l} \,,
\end{equation}
near the brane. This integrates to give
\begin{equation}
e^{\zeta/l}=1+\frac{t_0-t}{l} \,,
\end{equation}
for the past light-cone, where $t_0$ is the time today.
We want to compare this to (\ref{goodapprox}) to see when we can argue
that the approximation is good on causality grounds alone.
During a matter dominated era of the universe, the critical energy
density is given by
\begin{equation}
\rho=\frac{1}{6\pi G t^2}\,,
\end{equation}
and the bare tension on the brane is related to the AdS length-scale by
$\lambda=3/(4\pi Gl^2)$.  So, in the matter dominated era,
(\ref{goodapprox}) becomes
\begin{equation}
e^{\zeta/l}=\left(1+9\frac{t^2}{l^2}\right)^{1/2}\approx3\frac{t}{l}\,,
\end{equation}
which intersects the light-cone when $t \approx t_0/4$.
Of course, the approximation used in (\ref{nullDE}) is good near the
brane but breaks down as the null geodesic approaches the line
(\ref{goodapprox}) so this is only and order of magnitude estimate.
Nonetheless, this means that we cannot use a causality argument to
justify the approximation throughout the whole history --- we would
like to have obtained a time before the end of inflation as the
answer! It seems clear that some mode mixing will take place.

Finally, we should point out that the near-brane approximation is also
equivalent to an expansion in powers of $(Hl)^2$ (or
$\rho/\lambda$). Since it involves ignoring the second term in
(\ref{aexpr}), one could do this either because $\sinh(\zeta/l)$ is
small, or because $\rho/\lambda\propto (Hl)^2$ is small.

\subsection[Minkowski brane]{Minkowski brane}

For the RS case where the projected metric on the brane
is Minkowski, the functions $a$ and $n$ have the simple, separable form
\begin{equation}
a(\tau,\zeta)=n(\tau,\zeta)=e^{-\zeta/l}\,.
\end{equation}
This simplifies (\ref{Rtens2}), giving
\begin{equation}
\label{tensradevol}
\frac{1}{2}\ddot{h}+\frac{k^2}{2}h=
2\left(\frac{\Lambda}{3}+\frac{1}{l^2}\right)a^2h+\frac{1}{2}
\frac{\dd^2}{\dd\zeta^2}\left(a^2h\right)\,.
\end{equation}
Or, equivalently, from (\ref{tensorpertevol})
\begin{equation}
\ddot{h}+k^2h=e^{-2\zeta/l}\left({h}''
-\frac{4}{l}{h}'\right) \,.
\end{equation}
This equation has a constant solution, corresponding to a metric
perturbation proportional to $e^{-2\zeta/l}$, which is the same as the
solution found by Randall and Sundrum\cite{RS2}.
They interpreted this as a graviton zero-mode localised on
the brane, thus having the appearance of a four-dimensional
gravitational perturbation.

To find a set of basis solutions, we perform a separation of
variables, using the ansatz
\begin{equation}
h(\tau,\zeta)=\psi(\tau)\phi(\zeta)\,.
\end{equation}
The equation of motion (\ref{tensradevol}) then separates into
\begin{eqnarray}
&&\ddot \psi_m+\left(k^2 + m^2\right)\psi_m = 0 \,,\nonumber \\
&&\phi_m'' -\frac{4}{l}\phi'_m + m^2 e^{2\zeta/l} \phi_m=0\,,
\end{eqnarray}
where $m^2$ is a separation constant.
Because $m^2$ has dimensions of mass, modes with $m^2\ne0$ are
referred to as massive modes, but it should be noted that $m^2$ can be
negative.
The solutions are
\begin{eqnarray}
\psi_m(\tau)&=&c\cos\Big(\sqrt{m^2+k^2}\,\tau\Big)
+d\sin\Big(\sqrt{m^2+k^2}\,\tau\Big) \,,\\
\phi_0(\zeta)&=&C+De^{4\zeta/l}\,,\\
\phi_m(\zeta)&=&e^{2\zeta/l}\,Z_2\Big(mle^{\zeta/l}\Big)\,,
\end{eqnarray}
where $c$, $d$, $C$ and $D$ are constants of integration and
$Z_\nu$ represents a linear combination of Bessel functions of order
$\nu$.  (The boundary conditions will determine which particular
combination to choose.)
Randall and Sundrum \cite{RS2} also showed that, in the Newtonian limit, the
contribution from the massive modes is sub--dominant, and so gravity
has approximately four-dimensional behaviour.

\subsection[De Sitter era]{In a de Sitter era}

During an era of inflation in the primordial universe, the density of
matter is almost constant, an so the spacetime is approximately a de
Sitter universe.
For this reason it is useful to consider a brane-world where the brane
metric takes de Sitter form.
For a de Sitter brane, the matter density of the background solution
is a constant so scale factors $a$ and $n$ are separable functions.
We shall work in conformal time, where they have the functional form
\begin{equation}
a(\tau,\zeta)=n(\tau,\zeta)=a_0(\tau){\cal A}(\zeta) \,,
\end{equation}
where $a_0(\tau)=a(\tau,0)$ and ${\cal A}(\zeta)$ is given by
\begin{equation}
{\cal A}(\zeta)=\left(1+\frac{\rho}{2\lambda}\right)e^{-\zeta/l}
+\frac{\rho}{2\lambda}e^{\zeta/l}\,.
\end{equation}
The wave equation (\ref{tensorpertevol}) becomes
\begin{equation}
\ddot h+2\frac{\dot{a}_0}{a_0}\dot h+k^2 h 
= {\cal A}^2 h'' + 4 {\cal A}'{\cal A} h'\,.
\end{equation}
We can find a set of basis solutions to this by performing the same
separation of variables used in the RS case.
Writing $h=\psi(\tau)\phi(\zeta)$, the wave equation splits into two
parts:
\begin{eqnarray}
\label{psimassivemotion}
&&\ddot \psi_m + 2\frac{\dot{a_0}}{a_0} \dot \psi_m 
+\left(k^2 + m^2 a_0^2\right)\psi_m = 0 \,, \\
&&\phi_m'' + 4 \frac{{\cal A}'}{{\cal A}} \phi'_m +
\frac{m^2}{{\cal A}^2}\phi_m=0\,,
\end{eqnarray}
where $m^2$ is a separation constant as before.

In a de Sitter era $\rho$ is a constant so, in conformal time, we have
$a(\tau,0)=-(H_{\rm inf}\tau)^{-1}$ where $H_{\rm inf}$ is the energy scale for inflation, giving the evolution of the tensors as
\begin{eqnarray}
&&\ddot\psi_m-\frac{2}{\tau}\dot\psi_m+\left(k^2+
\frac{m^2}{H_{\rm inf}^2\tau^2}\right)\psi_m=0\,,\\
&&\phi_m''+ 4 \frac{{\cal A}'}{{\cal A}} \phi'_m+
\frac{m^2}{{\cal A}^2}\phi_m=0 \,.
\end{eqnarray}
The solutions for $\psi$ are
\begin{equation}
\psi=\tau^{3/2}\left(c\tau^\alpha+d\tau^{-\alpha}\right)\,,
\end{equation}
for $k=0$, and 
\begin{equation}
\psi=\tau^{3/2}\Big(cJ_\alpha(k\tau)+dY_\alpha(k\tau)\Big)\,,
\end{equation}
for $k\ne0$, where $\alpha=\sqrt{9-(4m^2/H_{\rm inf}^2)}/2$ and $c$ and $d$ are
constants of integration.
The equation for $\phi$ reduces to a hypergeometric equation, which we
can be solved exactly, but the solution is very complicated and not
particularly informative.
Let us state briefly the $m=0$ solution, which is
\begin{equation}
\phi_0=C+De^{-4\zeta/l}\,.
\end{equation}

This de Sitter case has been studied extensively in the literature.
It was found that the Kaluza--Klein modes are not continuous, but there 
is a mass gap between the zero mode and the first massive mode. If one requires that $\alpha$ is pure imaginary then $m>(3/2)H_{\rm inf}$. Above that mass 
the spectrum is continuous. 
In \cite{LMW}~it was argued that the tensor perturbation left
after an inflationary era would mainly be comprised of the zero-mode
solution, $h=\psi_0(\tau)\phi_0(\zeta)$, and this result was made more
quantitative in~\cite{Rubakov} where Bogoliubov coefficients were
calculated.
Thus, the perturbations produced from a period of inflation will be the
zero-mode, $h=\psi_0(\tau)\phi_0(\tau)$, to good approximation.
This has been extended to the case of any conformally flat brane
metric in~\cite{FrolovKofman}.

\subsection{In the radiation era}
\label{SecRadEra}

Let us now consider the evolution of the tensor perturbations in the 
radiation-dominated epoch.
As mentioned earlier, we cannot solve the equations exactly and so we
use the near-brane approximation.
As for the de Sitter era, we work in conformal time, where the the
scale factors $a$ and $n$ have the form
\begin{eqnarray}
\label{nearbranescalefactors}
a(\tau,\zeta) = n(\tau,\zeta) = a_0(\tau)e^{-\zeta/l}\,.
\end{eqnarray}
We are assuming that the density is small compared to the bare tension
of the brane in order for the near-brane solution to be a good
approximation, so we can approximate the Friedmann equation by the
usual, four-dimensional one
\begin{equation}
\label{4DF}
\left(\frac{\dd a}{\dd\tau}\right)^2=\frac{8\pi G}{3}\rho a^4 \,,
\end{equation}
which, if we choose the origin of time appropriately, has solution
$a=A\tau$ for some constant $A$.

Our near-brane approximation allows us to separate variables, so we
try to find solutions of the form $h=\psi(\tau)\phi(\zeta)$ as
before. 
The wave equation then splits into two equations of motion
\begin{eqnarray}
\label{psiradeq}
&&\ddot \psi_m + \frac{2}{\tau} \dot \psi_m +
\left[k^2 + m^2A^2\tau^2\right] \psi_m = 0 \,,\\
&&\phi_m''-\frac{4}{l}\phi_m'+m^2e^{2\zeta/l}\phi_m=0\,.
\label{phiradeq}
\end{eqnarray}
The solution of (\ref{phiradeq}) is independent of $k$ and is given by
\begin{eqnarray}
\label{phisoln0}
\phi_0(\zeta)&=&C+De^{4\zeta/l}\,,\\
\phi_m(\zeta)&=&e^{2\zeta/l}\, Z_2\Big(mle^{\zeta/l}\Big) \,,
\label{phisolnm}
\end{eqnarray}
where $C$ and $D$ are integration constants, and, as before, $Z_\nu$ represents a
linear combination of Bessel functions of order $\nu$ with the boundary
condition determining the particular combination.
Assuming we have no source for the tensor perturbations, the boundary
condition is simply $\phi'=0$, which gives us basis solutions
\begin{equation}
\label{radsolnBBC}
\phi_m(\zeta)=e^{2\zeta/l}\Bigg\{Y_1(ml)J_2\Big(mle^{\zeta/l}\Big)
-J_1(ml)Y_2\Big(mle^{\zeta/l}\Big)\Bigg\}\,,
\end{equation}
with $J_\nu$ and $Y_\nu$ being, respectively, Bessel functions of the
first and second kind.
When $k=0$, the solutions to (\ref{psiradeq}) are
\begin{eqnarray}
\label{psisolnk0}
\psi_0(\tau)&=&c+\frac{d}{\tau} \,, \\
\psi_m(\tau)&=&\tau^{-1/2}\,Z_{1/4}\left(\frac{1}{2}mA\tau^2\right)\,,
\end{eqnarray}
and when $k\ne0$ the solution for $m=0$ is
\begin{equation}
\psi_0(\tau)=c\,\frac{\sin{k\tau}}{\tau}+d\,\frac{\cos{k\tau}}{\tau} \,,
\end{equation}
where $c$ and $d$ are constants of integration.

For $m\ne0$, the solutions can be written in terms of parabolic
cylinder functions as follows.
Writing $\psi=y/\tau$ and changing the independent
variable to $t=k\tau$, (\ref{psiradeq}) becomes
\begin{equation}
\label{depcf}
\frac{\dd^2y}{\dd t^2}+\left(1+\frac{m^2A^2}{k^4}t^2\right)y=0\,,
\end{equation}
which is Weber's equation and has parabolic cylinder functions as
solutions~\cite{Buchholz,GradRyz} 
\begin{equation}
D_{\nu_m}\left(\pm(1+i)\sqrt{mA}\,k^{-1}t\right)\,,\qquad \mbox{where}
\quad\nu_m=\frac{k^2}{2imA}-\frac{1}{2}\,.
\end{equation}
We can use the asymptotic form
\begin{equation}
D_\nu(z)\sim e^{-z^2/4}z^\nu\,,\qquad\mbox{for}\quad |z|\gg1\,,\quad
|z|\gg|\nu|\,,\quad |\arg z|<3\pi/4\,,
\end{equation}
to approximate the solutions by
\begin{equation}
\label{asympcf}
y_m \sim \frac{e^{k^2\pi/(8mA)}}{(2mA)^{1/4}\sqrt{\tau}}\:\exp
\left\{-\frac{i}{2}mA\tau^2-\frac{ik^2}{2mA}\log\tau\right\}\,,
\qquad\mbox{as}\quad \tau\rightarrow\infty
\end{equation}
which is used in Appendix A to derive the orthogonality relation.
If $k=0$ this can seen to have the same form as (\ref{psisolnk0}) as $\tau\rightarrow\infty$.
The general solutions can then be written as a superposition of these
modes of the form
\begin{equation}
h=C^{+}_0\,\frac{\sin{k\tau}}{\tau}+C_0^{-}\,\frac{\cos{k\tau}}{\tau}
+\sum_{\pm}\int^\infty \frac{C^{\pm}_m}{\tau}
\:D_{\nu_m}\!\left(\pm(1+i)\sqrt{mA}\,\tau\right)
\left\{Y_1(ml)J_2\Big(mle^{\zeta/l}\Big)
-J_1(ml)Y_2\Big(mle^{\zeta/l}\Big)\right\}\,\dd m\,.
\end{equation}
The evolution of the zero-mode is exactly the same as for the usual,
four-dimensional cosmology based on General Relativity. For all 
values of $k$, the very heavy modes decay outside the horizon.

Since special functions of this nature are not always easy to
visualize, let us try to approximate to solution to (\ref{depcf})
using the WKB method.  First define the dimensionless parameter
\begin{equation}
\epsilon=\frac{k^2}{mA}
\end{equation}
The WKB approximation for $\epsilon$ small recovers the asymptotic
expansion given above.  When $\epsilon$ is large, we rescale the time
coordinate by $T = t/\epsilon = mA\tau/k$, so that (\ref{depcf}) is
rewritten as 
\begin{equation}
\frac{1}{\epsilon^2}\frac{\dd^2y}{\dd T^2}+\big(1+T^2\big)y=0
\end{equation}
We will use the formulation of the WKB approximation as described in
\cite{BenderOrszag}, where a small parameter $\de$ is introduced and
an asymptotic solution of the form
\begin{equation}
\label{WKBseries}
\exp\left\{\frac{1}{\de}\sum_{n=0}^\infty\de^nS_n(T)\right\}\,,
\end{equation}
is postulated and the $S_n$ are functions to be determined.
The parameter $\de$ will be determined in terms of $\epsilon$ by a
distinguished limit, i.e., the choice for which a non-trivial answer
is obtained.
Substituting (\ref{WKBseries}) into the differential equation gives
\begin{equation}
\label{asymexp}
\frac{1}{\epsilon^2\de^2}\left(\sum\de^n\dot{S}_n\right)^2
+\frac{1}{\epsilon^2\de}\sum\de^n\ddot{S}_n=-1-T^2
\end{equation}
The term proportional to $(\epsilon\de)^{-2}$ is the leading term on the
left and so must match the right hand side.  Hence, we see that
$\de=1/\epsilon$ is the distinguished limit and
\begin{equation}
\dot{S}_0=\pm i\sqrt{1+T^2}\,,\qquad
\end{equation}
The coefficients of the other powers of $\epsilon$ give differential
equations for the other $S_n$: for the next three, we have
\begin{equation}
\dot{S}_1=-\frac{T}{2(1+T^2)}\,,\qquad
\dot{S}_2=\pm i \frac{3T^2-2}{8(1+T^2)^{5/2}}\,,\qquad
\dot{S}_3=\frac{3T(2T^2-3)}{8(1+T^2)^4}\,.
\end{equation}
The solutions to these equations are
\begin{eqnarray}
S_0&=&\pm \frac{i}{2}\left(T\sqrt{1+T^2}+\text{arcsinh}(T)\right)\,,\\
S_1&=&-\frac{1}{4}\log(1+T^2)\,,\\
S_2&=&\mp \frac{i}{24}\frac{T(T^2+6)}{(1+T^2)^{3/2}}\,,\\
S_3&=&\frac{2-3T^2}{16(1+T^2)^3}\,,
\end{eqnarray}
which will give an approximation to $y_m$.  Note that constants of
integration have been omitted because they merely correspond to
multiplying the solution by a constant factor.

It is useful to consider the early and late time behaviour of this approximation.
When $T$ is large, the asymptotic expansion (\ref{asympcf}) is recovered.  
When $T$ is small, the $S_n$ are approximately
\begin{equation}
S_0\approx\pm iT\,,\qquad
S_1\approx-\frac{1}{4}T^2\,,\qquad
S_2\approx\mp \frac{1}{4}iT\,,\qquad
S_3\approx-\frac{9}{16}T^2\,,
\end{equation}
so the solution is
\begin{equation}
y_m\approx\exp\left[-\frac{T^2}{4}\left(1+\frac{9}{4\epsilon^2}\right)\right]
\exp\left[\pm iT\left(\epsilon-\frac{1}{4\epsilon}\right)\right]\,,
\end{equation}
which is a good approximation for $k\tau\ll\epsilon$.
We see that the solution which is bounded at the origin is
\begin{equation}
\psi_m\approx\frac{1}{T}\exp\left[-\frac{T^2}{4}\left(1+\frac{9}{4\epsilon^2}\right)\right]
\sin\left[T\left(\epsilon-\frac{1}{4\epsilon}\right)\right]\,,
\end{equation}
which is flat near the origin and then begins to decay.

\subsection{In the matter era}

We now move on to the final phase of the evolution of cosmological
perturbations when the universe is dominated by a pressureless fluid.
As with the radiation era, we approximate the solution near the brane,
where the scale factors $a$ and $n$ have the form
(\ref{nearbranescalefactors}).
Again, we are quite late in the history of the universe, so
(\ref{4DF}) is valid to good approximation and we have 
$a= A\tau^2$ if the origin of time is chosen appropriately.
Making the ansatz $h=\psi(\tau)\phi(\zeta)$,
the equations of motion separate once again and the equation for $\psi$ is
\begin{equation}
\label{psimattereq}
\ddot \psi_m + \frac{4}{\tau} \dot \psi_m +
\left[k^2 + m^2A^2\tau^4\right] \psi_m = 0 \,.
\end{equation}
The equation for $\phi$ and its solutions are the same as in the radiation era
(\ref{phiradeq}).
For $k=0$, we can solve (\ref{psimattereq}) exactly, giving
\begin{eqnarray}
\label{psimatterm0}
\psi_0(\tau)&=&c+\frac{d}{\tau^3}\,,\\
\psi_m(\tau)&=&\frac{c}{\tau^3}\cos\left(\frac{1}{3}mA\tau^3\right)+
\frac{d}{\tau^3}\sin\left(\frac{1}{3}mA\tau^3\right)\,.
\label{psimatterm}
\end{eqnarray}
For $k\ne0$, the zero-mode can be found exactly
\begin{equation}
\psi_0(\tau)=\frac{c}{\tau^3}\Big(k\tau\sin(k\tau)+\cos(k\tau)\Big)
+\frac{d}{\tau^3}\Big(k\tau\cos(k\tau)-\sin(k\tau)\Big)\,.
\end{equation}

We will study the modes with $k\ne0$ and
$m\ne0$ by constructing approximate solutions, which will be
appropriate in various limits. First, we will define the dimensionless
variable
\begin{equation}
\eps=\frac{k^3}{mA}\,.
\end{equation}
and two rescaled time variables $t=k\tau$ and $T=\eps^{-1/2}t=(mA/k)^{1/2}\tau$
which we will use in the limits $\eps\ll 1$ and $\eps\gg 1$
respectively. The equations of motion for $\psi$ are then 
\begin{equation}
\label{rescaledtensor1}
\eps^2\left(\frac{\dd^2\psi}{\dd t^2}+\frac{4}{t}\frac{\dd\psi}{\dd t}
+\psi\right)+t^4\psi=0\,,
\end{equation}
and 
\begin{equation}
\label{rescaledtensor2}
\eps^{-1}\left(\frac{\dd^2\psi}{\dd T^2}+\frac{4}{T}\frac{\dd\psi}{\dd T}\right)
+\big(1+T^4\big)\psi=0\,.
\end{equation}

\subsubsection*{Case $\eps\ll 1$:}
Let us consider first the case where $\eps$ is small and use
(\ref{rescaledtensor1}). We can then find a solution using the WKB
approximation as discussed in the previous section.
We try to find an asymptotic solution of the form (\ref{WKBseries}),
which, when substituted into (\ref{rescaledtensor1}) gives
\begin{equation}
\frac{\eps^2}{\de^2}\left(\sum\de^n\dot{S}_n\right)^2
+\frac{\eps^2}{\de}\sum\de^n\ddot{S}_n+\frac{4}{t}\frac{\eps^2}{\de}
\sum\de^n\dot{S}_n+\eps^2=-t^4\,.
\end{equation}
We see that the distinguished limit is when $\eps=\de$ and that the
$S_n$ are given by the differential equations
\begin{eqnarray}
&&\dot{S}_0^2=-t^4\,,\\
&&2\dot{S}_0\dot{S}_1+\ddot{S}_0+\frac{4}{t}\dot{S}_0=0\,,\\
&&2\dot{S}_0\dot{S}_2+\dot{S}_1^2+\ddot{S}_1+\frac{4}{t}\dot{S}_1+1=0\,,\\
&&2\dot{S}_0\dot{S}_n+\sum_{j=1}^{n-1}\dot{S}_j\dot{S}_{n-j}+\ddot{S}_{n-1}
+\frac{4}{t}\dot{S}_{n-1}=0\,,
\end{eqnarray}
for the first four terms.  Solving for these, we find
\begin{equation}
S_0=\pm\frac{1}{3}it^3\,,\qquad S_1=\log (t^{-3})\,,\qquad
S_2=\mp \frac{i}{2t} \,,\qquad S_3=-\frac{5}{t^6}\,.
\end{equation}
Substituting the first three of these into (\ref{WKBseries}) along with the expressions for
$\eps$ and $t$, the solution with the first correction term is
\begin{equation}
\psi=\frac{c}{\tau^3}\,e^{-5/(m^2A^2\tau^6)}\,\cos\left(\frac{1}{3}mA\tau^3-\frac{k^2}{2mA\tau}\right)
+\frac{d}{\tau^3}\,e^{-5/(m^2A^2\tau^6)}\,\sin\left(\frac{1}{3}mA\tau^3-\frac{k^2}{2mA\tau}\right)\,.
\end{equation}
This solution is good for small $k$ and is progressively better for
larger values of $\tau$. We see that in the limit $k\rightarrow 0$
this solution reverts to the form (\ref{psimatterm}) as one would
expect.

\subsubsection*{Case $\eps\gg1$:}
Now consider the opposite limit where $\eps$ is large using (\ref{rescaledtensor2}) which is in a form suitable for using the WKB method as formulated
above.
We get $\de=\eps^{-1/2}$ and the following differential equations,
\begin{eqnarray}
\label{Sde0}
&&\dot{S}_0=\pm i\sqrt{1+T^4}\,,\\
&&2\dot{S}_0\dot{S}_1+\ddot{S}_0+4T^{-1}\dot{S}_0=0\,,\\
&&2\dot{S}_0\dot{S}_2+\dot{S}_1^2+\ddot{S}_1+4T^{-1}\dot{S}_1=0\,,\\
&&2\dot{S}_0\dot{S}_3+2\dot{S}_1\dot{S}_2+\ddot{S}_2+4T^{-1}\dot{S}_2=0\,,
\label{Sde3}
\end{eqnarray}
for the first four terms; the last three equations simplify to
\begin{equation}
\dot{S}_1=-\frac{T^3}{1+T^4}-\frac{2}{T}\,,\qquad
\pm\dot{S}_2=\frac{7T^4+2}{2iT^2\big(1+T^4\big)^{5/2}}\,,\qquad
\dot S_3=-\frac{35T^8+7T^4+2}{T^3(1+T^4)^4}\,.
\end{equation}
These have solutions
\begin{eqnarray}
\pm S_0&=&\frac{i}{3}T\sqrt{1+T^4}+\frac{2}{3}e^{i\pi/4}
F\left(e^{i\pi/4}T,i\right)\,,\\
S_1&=&-\log\Bigg(T^2\big(1+T^4\big)^{1/4}\Bigg)\,,\\
\pm S_2&=&\frac{i}{T}\sqrt{1+T^4}-\frac{iT^3}{8\sqrt{1+T^4}}
-\frac{5iT^3}{12\big(1+T^4\big)^{3/2}}
+\frac{7}{8}e^{-i\pi/4}\Bigg[ F\left(e^{i\pi/4}T,i\right)
-E\left(e^{i\pi/4}T,i\right) \Bigg]\,,\\
S_3&=&\frac{7T^4+2}{2T^2(1+T^4)^3}\,,
\end{eqnarray}
where
\begin{equation}
F\left(e^{i\pi/4}T,i\right)=\int_0^T\frac{\dd x}{\sqrt{1+x^4}}\,,\qquad
E\left(e^{i\pi/4}T,i\right)=\int_0^T\sqrt{\frac{1+ix^2}{1-ix^2}}\,\dd x
\end{equation}
are, respectively, elliptic integrals of the first and second kinds.
These expressions are quite complicated, so let us consider the two limits where
$T$ is small or large.

When $T$ is small, the differential equations (\ref{Sde0}-\ref{Sde3}) are
\begin{equation}
\dot{S}_0\approx\pm i\,,\qquad
\dot{S}_1\approx-\frac{2}{T}\,,\qquad
\dot{S}_2\approx\mp\frac{i}{T^2}\,,\qquad
\dot{S}_3\approx-\frac{2}{T^3}\,,
\end{equation}
so that the solution is approximately
\begin{equation}
\psi(T)\approx\frac{e^{1/(\eps T^2)}}{T^2}
\exp\left[\pm i\left(\sqrt{\eps}\,T+\frac{1}{\sqrt{\eps}\,T}\right)\right]
\propto\frac{e^{1/(k^2\tau^2)}}{\tau^2}
\exp\left[\pm i\left(k\tau+\frac{1}{k\tau}\right)\right]
\end{equation}
which should be a good approximation if $1 \ll k\tau \ll \sqrt\eps$.
This approximate solution for small $T$ does not depend on $m$, which
is a consequence of the fact that the term involving $m$ is the
equation of motion has a $T^4$ dependence.
When $T$ is large equations (\ref{Sde0}-\ref{Sde3}) become
\begin{equation}
\dot{S}_0\sim\pm iT^2\,,\qquad
\dot{S}_1\sim-\frac{3}{T}\,,\qquad
\dot{S}_2\sim\mp\frac{7i}{2T^8}\,,\qquad
\dot{S}_3\sim-\frac{35}{T^{11}}\,,
\end{equation}
giving us
\begin{equation}
\psi(T)\approx\frac{e^{7/(2\eps T^{10})}}{T^3}
\exp\left[\pm i\left(\frac{1}{3}\sqrt{\eps}\,T^3+\frac{1}{2\sqrt{\eps}\,T^7}\right)\right]
\propto\frac{1}{\tau^3}\exp\left(\frac{7k^2}{2m^4A^4\tau^{10}}\right)
\exp\left[\pm i\left(\frac{1}{3}mA\tau^3+\frac{k^2}{2m^3A^3\tau^7}\right)\right]
\end{equation}
valid for $k\tau \gg \sqrt{\eps}$.

\section{Imposing the initial conditions}
\label{secIC}
We are considering tensor perturbations here and since the matter on
the brane has no tensor perturbation early in the universe, there is
no matter source for these perturbations in the metric.
So the boundary condition at the brane is $h'=0$.
The $\zeta$ dependence of the solution near the brane is the same in
all eras, and its evolution is given by (\ref{phiradeq}), with
solutions given in (\ref{phisoln0}) and (\ref{phisolnm}).
The derivative of $\phi_m(\zeta)$ is
\begin{equation}
\phi'_m(\zeta)=m e^{3\zeta/l} Z_1\big(mle^{\zeta/l}\big)\,,
\end{equation}
so, if the junction condition is imposed, the massive modes are
\begin{equation}
\phi_m(\zeta)=e^{2\zeta/l} \left( Y_1(ml)J_2\big(mle^{\zeta/l}\big)
-J_1(ml)Y_2\big(mle^{\zeta/l}\big) \right)\,,
\end{equation}
for which
\begin{equation}
\phi'_m(\zeta)=m e^{3\zeta/l} \left( Y_1(ml)J_1\big(mle^{\zeta/l}\big)
-J_1(ml)Y_1\big(mle^{\zeta/l}\big) \right)\,.
\end{equation}
The zero-mode is simple a constant
\begin{equation}
\phi_0(\zeta)=C\,,
\end{equation}
when the boundary condition is imposed.

Let us now try to impose initial conditions at some initial time,
$\tau_i$, which will, in practice, be the transition between two
eras.
This initial data must satisfy the boundary condition and will be
expressible as a superposition of the mode solutions $\phi_m(\zeta)$.
This superposition is most easily calculated by considering the $\zeta$
derivative of the initial data.
So let us write
\begin{equation}
\label{comb}
h'(\tau_i,\zeta)=\int_0^\infty A(m)\phi'_m(\zeta)\, \dd m
=\int_0^\infty m A(m) e^{3\zeta/l} C_1\big(m ; le^{\zeta/l}\big) \,\dd m\,,
\end{equation}
where $C_\nu(m;z)$ is the cylinder function
\begin{equation}
C_\nu(m;z)= Y_1(ml)J_\nu(mz)-J_1(ml)Y_\nu(mz)\,.
\end{equation}
We can deduce the $A(m)$ from the orthogonality relation given in the
appendix.

The contribution from the zero-mode is just a constant, and this is,
of course, the bit that is undetermined by considering $h'$.
So we have a prescription for imposing any given initial conditions.
The prescription given is somewhat academic, at least for the
application under consideration here, because the spatial parts of the
mode functions, $\phi_m(\zeta)$, are the same in all eras, so the
spectrum of modes present at the beginning of an era will be the same
as the spectrum at the end of the previous era.
In particular if inflation produces a perturbation which is well
approximated by the zero-mode, as indeed it does according
to~\cite{Rubakov,LMW}, then only the zero-mode is present during
subsequent eras too.
Thus the evolution of tensor perturbations is the same as in the
standard, four-dimensional case, and hence the contribution towards
CMB anisotropies is the same.
The absence of sources is responsible for this similarity, tensor
matter perturbations on the brane would source the massive modes and
give an answer different from the standard cosmology.
The analysis presented here should extend to the scalar perturbations,
although there are some additional technical difficulties.
The scalar perturbations are not only the dominant contributions to
the CMB anisotropies but are sourced by scalar matter perturbations so
there will very probably be some signature of the RS model detectable
in the CMB power spectrum.

\section{Discussion and Conclusions}
\label{conc}

In this article we have discussed some aspects of tensor perturbations 
in the Randall--Sundrum brane-world. The problem can be formulated in 
several coordinate systems, each of which has its shortcomings. 
In GN coordinates, the brane is static and the boundary 
condition can easily be imposed. However, the equations of motion for 
tensor perturbations are not separable. By formulating the problem in 
static bulk coordinates in which the brane moves, the bulk equation
can be solved but the boundary conditions are difficult to impose on 
the (moving) brane. Therefore, an approximation scheme is necessary to
make progress in both systems. We have presented solutions in the  
de Sitter era, the radiation and matter dominated era using the near
brane approximation first discussed in \cite{andyproc,andythesis}.  

The near-brane approximation is akin to an adiabatic
treatment of the motion of the brane in that the warp factor
$a(\tau,\zeta)$ has the same $\zeta$ dependence as in the standard
Randall--Sundrum picture for a Minkowski brane, but is multiplied by
the cosmological scale factor. 
In this lowest-order approximation, independent modes  persist
throughout the radiation and matter eras. If this approximation is
universally valid, which we have argued against on causality grounds,
and the initial spectrum includes only the zero mode created during
inflation then the observed  tensor spectrum will be exactly that of
the standard cosmological model. 

The zero mode is equivalent to the standard case of General
Relativity, but the massive modes have some novel features. In
particular they can be seen to decay and oscillate  outside the
horizon, whereas standard massless gravitons only begin to oscillate
at horizon crossing. One important consequence of this is that initial
power spectrum of fluctuations in the massive modes cannot be scale
invariant as $k\rightarrow 0$. There must exist a cut-off in the
spectrum at very small values of $k$. If one thinks of the
perturbations as being particles created quantum mechanically during
inflation then one can create those with $k\ll m$ just on energetic
grounds. Moreover, if massive modes exist and they decay outside the
horizon, the photon quadrupole will increase without the need for
Thomson scattering since it will be sourced by the decay of the
super-horizon gravitational potential. This could lead to a
significant polarization signature in the angular power spectrum on
large angular scales. 

It is worth noting that the equations of motion for the massive modes
which we have solved here are equivalent to those of ordinary massless
gravity with a mass term added to the Lagrangian in an \emph{ad hoc}
way via an addition term of the form  
\begin{equation}
{\cal L}_{\rm mass}=-m^2\left(h_{\mu\nu}h^{\mu\nu}-h^2\right)\,,
\end{equation}
where $m$ is the graviton mass and $h_{\mu\nu}$ is the 4D metric perturbation.
Although such a theory is believed to suffer from the van Dam-Veltman
discontinuity~\cite{vanD} in the graviton propagator in going from
$m=0$ to $m>0$, this theorem will not apply in the RSII case where
there is likely to be a spectrum of modes including the zero mode.

It should be possible to incorporate non-adiabatic effects, by taking
into account the non-separability of the equations. We anticipate that
this will lead to mixing of the near-brane modes discussed in this
paper, as considered in \cite{easther,HKT}. This is likely to lead to
non-trivial time evolution of the total perturbation amplitude and
interesting cosmological signatures. 
The solutions found in this paper can  be used to gain insights into
the full problem. The non-linearity in the wave equation induced
next-order correction to the expressions (\ref{sfa}) and (\ref{sfn})
for the scale factors $a(\tau,\zeta)$ and $n(\tau,\zeta)$ is the
origin of the mode mixing phenomena in \cite{HKT}.

The inclusion of a tensor matter component on the brane will source
massive modes. The main source of this will be the photon quadrupole
which develops late in the history of the universe. This issue will
even more important in the case of scalar perturbations since the
perturbations in the density, pressure and velocity will contribute to
this. Some of the methods discussed here should be of use in taking this
effect into account.

\vspace{0.5cm}

\begin{acknowledgments}
RAB and CvdB are supported by PPARC; AM is supported by Emmanuel
College, Cambridge.
We acknowledge useful conversations with Martin Bucher and Toby Wiseman. 
\end{acknowledgments}

\appendix*
\section{Orthogonality relations}
Sturm--Liouville theory tells us that our basis functions will be
orthogonal, but we need to know the normalization.
We will derive the orthogonality here explicitly for the Bessel and
parabolic cylinder functions used and determine the coefficient of the
delta function in the orthogonality relation.

There is a standard formula for Bessel functions which tells us that
\begin{equation}
\int x D_\nu(mx) \tilde{D}_\nu(nx) \,\dd x = \frac{x}{m^2-n^2}
\Big( m D_{\nu+1}(mx) \tilde{D}_\nu(nx) -
n D_\nu(mx) \tilde{D}_{\nu+1}(nx) \Big) \,,
\end{equation}
where $D_\nu$ and $\tilde{D}_\nu$ are any two Bessel functions of
order $\nu$ and $m \ne n$.
This can be derived by an argument given in~\cite{Watson}.
If we consider two functions $g_1(x)$ and $g_2(x)$ which satisfy the differential
equations $g_n''+P_n(x)g_n=0$  for some functions $P_n$, then 
\begin{equation}
\int (P_1-P_2)g_1g_2\,\dd x=g_1g_2'-g_1'g_2\,.
\end{equation}
Applying this to Bessel's equation, we see that
\begin{equation}
\int^\infty_{x_0} x D_\nu(mx) \tilde{D}_\nu(nx) \,\dd x \propto \de(m-n)\,,
\end{equation}
if $D_\nu(nx)$ and $\tilde{D}_\nu(nx)$ are chosen to vanish, or have
vanishing derivatives, on the boundary.
We are interested in the specific case
\begin{equation}
D_\nu(mx)=C_\mu(m;x)=Y_1(ml)J_\nu(mx)-J_1(ml)Y_\nu(mx)\,,
\end{equation}
with $\nu=1$, and
\begin{equation}
C_\mu(m;x) \sim \sqrt{\frac{2}{\pi mx}} \Big(Y_1(ml)^2+J_1(ml)^2\Big)
\sin\big(mx+\text{phase angle}\big)\,,
\end{equation}
asymptotically, when $x$ is large.
So the constant of proportionality can be deduced~\cite{Jackson} to be
\begin{equation}
\int^\infty_{x_0} D_\nu(mx) \tilde{D}_\nu(nx) \,\dd x
=\frac{1}{m}\Big(Y_1(ml)^2+J_1(ml)^2\Big)\,\de(m-n)\,.
\end{equation}

For the time-dependent parts of the mode solutions,
$\psi_m(\tau)=\tau^{-1}y_m(\tau)$, we can construct a similar
orthogonality relation
The $y_m$ satisfy the differential equation (\ref{depcf})
\begin{equation}
\frac{\dd^2y_m}{\dd t^2}+\left(1+\frac{m^2A^2}{k^4}t^2\right)y_m=0\,,
\end{equation}
so applying the same argument as before gives
\begin{equation}
\frac{A^2}{k^4}\big(m^2-n^2\big)\int t^2 y_m y_n \,\dd t
=y_m y'_n - y'_m y_n \,,
\end{equation}
where the r.h.s.\ will be zero due to the boundary conditions.
To evaluate $\int t^2y_m y_n \dd t$ when $m=n$, we can use the
asymptotic form (\ref{asympcf})
\begin{equation}
y_m \sim \frac{e^{k^2\pi/8mA}}{(2mA)^{1/4}\sqrt{\tau}}\:\exp
\left\{-\frac{i}{2}mA\tau^2-\frac{ik^2}{2mA}\log\tau+i\theta_m\right\}\,,
\end{equation}
to write the integrand as
\begin{equation}
\frac{e^{k^2\pi/8mA}e^{k^2\pi/8nA}}{\sqrt{2A}(mn)^{1/4}}\,kt
\:\exp\left\{-\frac{i}{2}(m+n)At^2-\frac{ik^2}{2mA}\log(k^{-1}t)
-\frac{ik^2}{2nA}\log(k^{-1}t)\right\}
\end{equation}
Neglecting the log terms in the exponential and making the change of
variables $x=At^2$
\begin{equation}
\int t^2 y_m y_n \,\dd t \sim
\frac{ke^{k^2\pi/8mA}e^{k^2\pi/8nA}}{(2A)^{3/2}(mn)^{1/4}}
\:\int e^{-imx/2}e^{-inx/2}\,\dd x
\end{equation}
which is asymptotically the same as $\int x J_\mu(mx/2)J_\mu(nx/2)$,
allowing us to calculate the coefficient of the delta function.

\end{document}